# Roles of discount rate, risk premium, and device performance in estimating the cost of energy for photovoltaics

Sergei Manzhos[1]

Department of Mechanical Engineering, Faculty of Engineering, National University of Singapore, Block EA #07-08, 9 Engineering Drive 1, Singapore 117576

**Abstract**

We show that different rates should be used for borrowing and discount rates, and that the risk-free rate should be used for discounting when assessing and comparing the cost of energy accross diffferent producers and technologies, on the example of photovoltaics. Recent quantitative models using the same rate for borrowing and discounting lead to an underestimation of the cost for risky borrowers and to distorted sensitivities of the cost to financial and non-financial factors. Specifically, it is shown that they may lead to gross underestimation of the importance of solar-to-electricity conversion efficiency. The importance of device efficiency is re-established under the treatment of the discount rate proposed here. The effects on the cost of energy of the installation efficiency and degradation rate, on the discount rate and risk premium as well as on the project lifetime are estimated.

---

[1] Email: mpemanzh@nus.edu.sg

Phone: +65 6516 4605

Realistic estimates of the costs of and cash flow from photovoltaic (PV) installations are necessary to guide rational resource allocation and to create a viable PV industry (OOPEC, 2007). They are also important to help establish research priorities. This is all the more important, as achievable costs are still significantly higher than those for hydro, nuclear, coal, and gas powered electricity (Azzopardi, 2011). The advent of new drilling technologies and the concomitant collapse of natural gas prices in North America in recent years (IEA, 2011), the achievement of economical recoverability of vast deposits of oil in Canada (Canada National Energy Board, 2004) and of new deposits in South and Central America (USGS, 2000; BP, 2011), coming at a time of strained public finances around the developed world set the bar for PV higher today than was the case just a few years ago, as policy-makers may become less willing to renew or extend subsidies. It is therefore critical to develop quantitative methods for realistic assessment of costs. Recently, progress has been made in establishing a quantitative methodology to estimate such costs (Azzopardi, 2011; Darling, 2011; Branker, 2011). Uncertainties surrounding the values of input parameters can be taken into account via Monte Carlo simulations sampling likely distributions of parameter values (Darling 2011). With such models, it is possible today to obtain ballpark estimates of the costs and their breakdown by component, as well as the sensitivity of the overall electricity cost from a PV installation to various physical and economic factors.

Specifically, in Darling (2011), levelized cost of energy (LCOE) from photovoltaic installations was computed based on a number of assumptions about insolation, solar cell efficiency and its temporal degradation, as well as prevailing financial variables such as capital costs, and tax rates or subsidies. This analysis assumed a mean power conversion efficiency of 16%, i.e. relevant for conventional solid state solar cells. The authors arrive at a cost of about 7-10 ¢/kWh. A most interesting result comes from the analysis of sensitivity of the LCOE to various factors. For example, the rank correlation sensitivity is by far the largest to the real discount rate, at 0.9, while factors related to the ability of the device to convert solar energy to electricity are much less important: -0.3 for conversion efficiency, 0.2 for system degradation, and -0.1 for insolation. The relative importance of fixed operation and maintenance was the lowest at below 0.1 (see Fig. 7 of Darling (2011)). The analysis was based on the following equation

$$LCOE = \frac{PCI - \sum_{n=1}^{N}\frac{DEP+I}{(1+DR)^n}TR + \sum_{n=1}^{N}\frac{LP}{(1+DR)^n} + \sum_{n=1}^{N}\frac{AO(1-TR)}{(1+DR)^n} - \frac{RV}{(1+DR)^n}}{\sum_{n=1}^{N}\frac{Initial\ kWh \times (1-SDR)^n}{(1+DR)^n}}, \quad (1)$$

where *PCI* is the project cost, *DEP* is depreciation, *I* is interest paid, *LP* is loan payment, *AO* are annual outlays (cost of operation), *TR* is the tax rate, *SDR* is system degradation rate, and *DR* is the discount rate. *Initial kWh* is output in year 1. The installation is assumed to operate for *N* years after which it has a residual value *RV* (Darling, 2011). A similar equation was used in Branker (2011).

In Azzopardi (2011), PV module costs and electricity cost per kWh were estimated for organic-based photovoltaics, OPV (assumed cell efficiency of 3% and 7%). The module cost was estimated to be between 63 and 192 €/m$^2$ with material costs accounting for 65 to 81% of the total. The LCOE was estimated to lie between 0.19 and 0.50 €/kWh for installations using 7% efficient cells. The cost per kWh was linear with respect to hardware costs and inversely proportional to the insolation and conversion efficiency (see Fig. 7 of Azzopardi (2011), to compare to Fig. 7 of Darling (2011)). These results were computed by expressing Life Cycle Investment Cost (LCIC, the equivalent of the numerator in Eq. (1)) of the module as

$$LCIC = \sum_{n=0}^{int(N/L_m)} \frac{C_{BOM}}{(1+DR)^{nL_m}} - \frac{C_{BOM}}{(1+DR)^N} E_{fr}^{rem}, \quad (2)$$

where $C_{BOM}$ is the present cost of the module, $E_{fr}^{rem}$ is the fraction of energy remaining, and int() takes the integer part.

A common feature of both these analyses is the use of the same value for the borrowing rate and the discount rate *DR*. The real rate was assumed to be 7% in Azzopardi (2011), and it was assumed to be distributed around 8% in Darling (2011), which corresponds to prevailing financing conditions for PV installations at the time of writing. Considering long plant and loan lifetimes of 25 and 30 years assumed in Azzopardi (2011) and Darling (2011), respectively, discounting at these rates will have a dramatic effect on the cost and sensitivity analysis.

First, we argue that if LCOE is defined as the average cost per kWh of *physical* output (as implied by Eq. (1)) over *N* years in present dollars, discounting should not be applied to the electricity produced, the denominator of Eq. (1). It is the present monetary values of (future) electricity sales and of the cost incurred to produce a kWh of electricity in the future that are obtained by discounting, not the physical output. In the event that the price per kWh of future sales is known, the denominator of Eq. (1) can be multiplied by it and also by the discount factor to perform a cost-benefit analysis. In the present case, however, we are dealing with the average cost per kWh in present-day dollars, and the present value of a future kWh is only defined in monetary terms. The application of the discount factor in the denominator of Eq. (1) leads to underestimation of physical electricity output over the life cycle (and a significant one given the rates used) and therefore to an overestimation of the cost. *DR* in the denominator of Eq. (1) effectively destroys much of the future output capacity and is equivalent to a higher *SDR*. This is one reason why *DR* was found to be by far the most important factor contributing to the cost (Darling, 2011). Discounting was not applied to physical output in (Azzopardi, 2011), with the result that the PV efficiency and other physical performance parameters took a more prominent role (see Fig. 7 therein). This also means that the rank correlation sensitivities to *SDR* and to *DR* in Fig. 7 of Darling (2011) are intermixed and do not provide a realistic picture of the influence of either parameter. We now argue that *DR* should not be set to the financing rate. Without a loss of generality of the argument about appropriate rates and to simplify the equations, we assume that the PV installation is financed by emitting a zero-coupon bond at rate *r*, due in *N* years. Further, we assume that the installation is not maintained or, equivalently, that maintenance and operation costs for every year were pre-funded by buying bond strips maturing in respective years and any *forces majeures* were insured against at the time of entering service, and these costs are added to *PCI*. We assume that the plant is discarded after *N* years (or, equivalently, that the cost of decommissioning minus *RV* is funded by buying a bond maturing in *N* year whose price is included into *PCI*). Nominal rates and no tax incentives are considered. Then Eq. (1) becomes

$$LCOE = \frac{PCI(1+r)^N}{(1+DR)^N} \times \frac{1}{Initial\ kWh \sum_{n=1}^{N}(1-SDR)^n}, \qquad (3)$$

where *Initial kWh* is assumed to be proportional to both the power conversion efficiency $\eta$ and to insolation *s*. In general, $\eta$ is a function of *s*, and this is a simplifying assumption that does however not influence the analysis of the role of financial variables. This assumption is also implied in Azzopardi (2011) and Darling (2011). Different sensitivities to $\eta$ and *s* computed in Darling (2011) are due to the non-linear nature of the Monte Carlo simulation which sampled them from different normal distributions. Here, we focus on the choice of the discount rate, and it is sufficient to use linear sensitivities for this purpose. It is obvious that Eq. (3) results in the relative sensitivity of 1 of the cost to $\eta$ and *s*, similar to Azzopardi (2011).

The first fraction in Eq. (3) is the present cost of the future bond repayment, which is also the cost of all electricity produced over the life cycle. If we set, as was done in Azzopardi (2011) and Darling (2011), *r=DR*, then this cost is independent of *r*, i.e. of the creditworthiness of the borrower. If this sounds implausible, it is because it's wrong. The absurdity of setting *DR* equal to the borrowing rate can also be seen from Eq. (1), where a higher borrowing rate decreases the contribution of various expenses (such as *AO* and *LP*) incurred over the life cycle to the total cost in the numerator for a riskier borrower due to a simultaneously higher *DR*.

The credit risk of the borrower is in the spread of interest payment *I* in Eq. (1) or borrowing rate *r* in Eq. (3) over the risk-free rate $r_0$, whereas the discount rate should reflect the present value of future money for players in the relevant market (Hull, 2009). Even if subjective present value of future money will differ for different agents (see, for example, a recent review, Branker (2011)), here we consider the objective cost of production which must be comparable among different producers and technologies. Indeed, the studies like those of (Azzopardi, 2011; Branker, 2011; Darling, 2011) are only useful if they are able to describe economic viability of solar-to-electricity conversion technologies, which is defined only in comparison with other technologies where borrowing rates and discount rates used in their respective actuarial calculations are widely different from and independent of those used by the solar industry (McIlveen-Wright, 2011). This argues for the use of the same *DR* for all producers when costs are compared.

It can further be argued that *DR* should tend to converge to the risk free rate. Indeed, any other rate would lead to an arbitrage opportunity (Bjork, 2009). Suppose a market agent can borrow at a rate *r<DR*. In that case, he can increase his present net value by long-term borrowing of large amounts of money, as the present value of future liabilities will be made arbitrarily small via $(1+r)^N / (1+DR)^N \to 0$. This will continue until the resulting increased credit risk leads to $r=DR$. The lowest $r$ among market participants is that of a large liquid sovereign who controls its own monetary policy. Therefore $DR=r_0$, the risk-free rate. In other words, the assumption $r=DR$ made in (Azzopardi, 2011; Darling, 2011) is only valid for risk-free borrowers (who can invest risk-free and borrow at the same rate). Alternatively, one can argue that the cost of bringing consumption (i.e. *PCI* of Eqs. 1, 3) forward in time should be the same as the benefit of deferred consumption. The latter is described by $r_0$, not by borrowing or "preferred" discount rates for particular agents or activities. Another way to rationalize the use of $r_0$ is to remember that the discount rate must include perceived risk used to convert future payments or receipts to present value. While the future cash flow from electricity sales can be uncertain, the obligation of loan repayment becomes certainty as soon as the loan is taken, and these outlays therefore should be discounted at the risk-free rate.

Consequently, *DR* of Eq. (3) should be equal to the rate on a zero-coupon Treasury note maturing in *N* years, and in Eq. (1), different *DR* should be used for each *n*, taken from the term structure of the Treasury market of the relevant sovereign.

We now present an analysis of the dependence of the nominal *LCOE* on *N*, *SDR*, *DR*, and *r* based on Eq. (3) around prevailing rates at the time of writing. First, we estimate the behavior of *LCOE* as a function of *N* and for $SDR=0.6\%$ (Moore, 2008). The increase of lifetime of OPV modules was identified in Azzopardi (2011) as essential to the achievement of economic viability. The cost was found to drop with *N* and largely to level off after 15 years (Azzopardi, 2011). In our analysis, we used a model term structure of the US Treasury market shown in Fig. 1 to obtain *DR* at any *N* (we neglect the small difference in yields between zero-coupon and conventional bonds which is unessential for the present analysis). The spread of *r* over *DR* was held fixed at 5% or 8% (for a maximum borrowing rate of about 11% at 30 years). In reality, the credit risk is term-dependent with the possibility of longer dated maturities having a wider or narrower spread over $r_0$. Our assumption is necessary considering the generality of the argument, and it does not affect qualitatively the conclusions.

The resulting dependence of the cost factor

$$cost\ factor = \frac{(1+r)^N}{(1+DR)^N \sum_{n=1}^{N}(1-SDR)^n} \quad (4)$$

on $N$ is also shown in Fig. 1. The cost factor bottoms at $N_{min}$ which depends on the credit spread and increases afterwards the faster the higher the $r$. This is the true influence of the borrowing rate: it limits both the minimum cost and the optimal $N$ due to escalating interest cost. In Fig. 7 of Azzopardi (2011) (analogous to our Fig. 1), the extent of the drop from $N=1$ to $N_{min}$ is slightly larger, and the cost keeps decreasing with $N$ seemingly forever. Here, for simplicity and as in Darling (2011), we assumed that the duration of the loan is equal to the lifetime of the installation, which does not have to be the case.

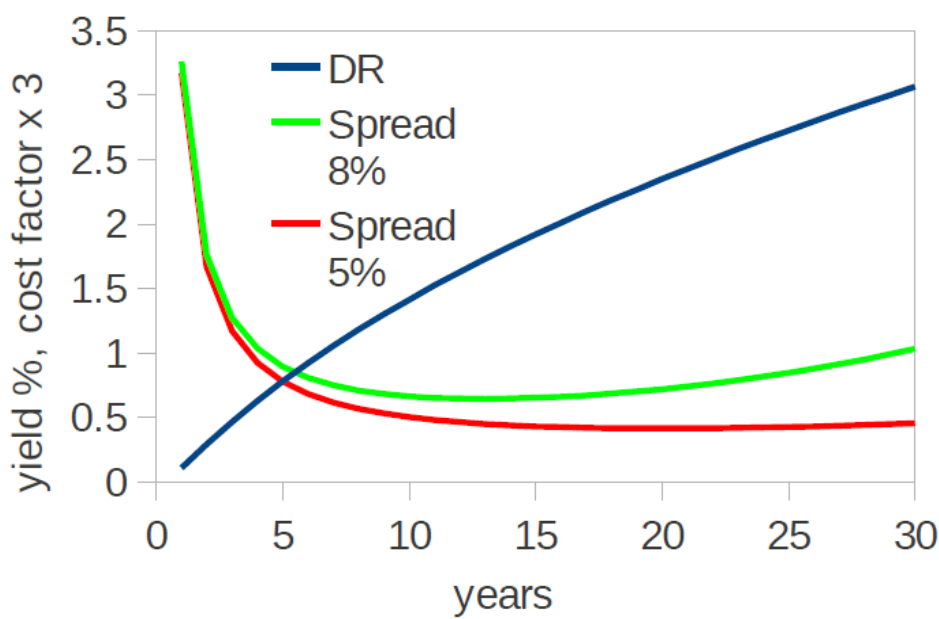

**Fig. 1.** A model US Treasury yield curve (blue) and the cost factor, Eq. (4) for the spread over the risk-free rate of 5% (red) and 8% (green). The US Treasury yield curve was taken from www.finance.yahoo.com. The equation we used is $yield\% = 0.0034(1.2892\ln(years) + 2.7061)^{3.473272}$, which approximates the yield curve as of October 3, 2011.

In Fig. 2, we plot the relative cost factor for different $SDR$, for different $DR$ at a fixed spread $r$-$DR$=5%, and for different credit spreads at $DR$=3%, for $N$=30 years. The curves are normalized to the cost at $r$-$DR$=5%, $SDR$=0.6%, and $DR$=3%. The system degradation rate has a rather mild effect on cost, doubling it as $SDR$ increases from 0.6% to about 5.6% per year, where at the end of service the installation will have only about 18% of its initial output. An increase of $DR$ has the effect of slightly decreasing the relative cost. This is because an increase of *DR* lowers the present cost of future obligations even as *r* grows as long as the credit risk (*r-DR*) does not increase. This $DR=r_0$ is determined by macroeconomic conditions and monetary policy and is not influenced by the borrower or by a particular industry. Both *SDR* and *DR* are less important than the conversion efficiency or insolation, which influence the relative cost per kWh as 1:1 (i.e. $\Delta LCOE/LCOE : \Delta(\eta,s)/(\eta,s) = 1$).

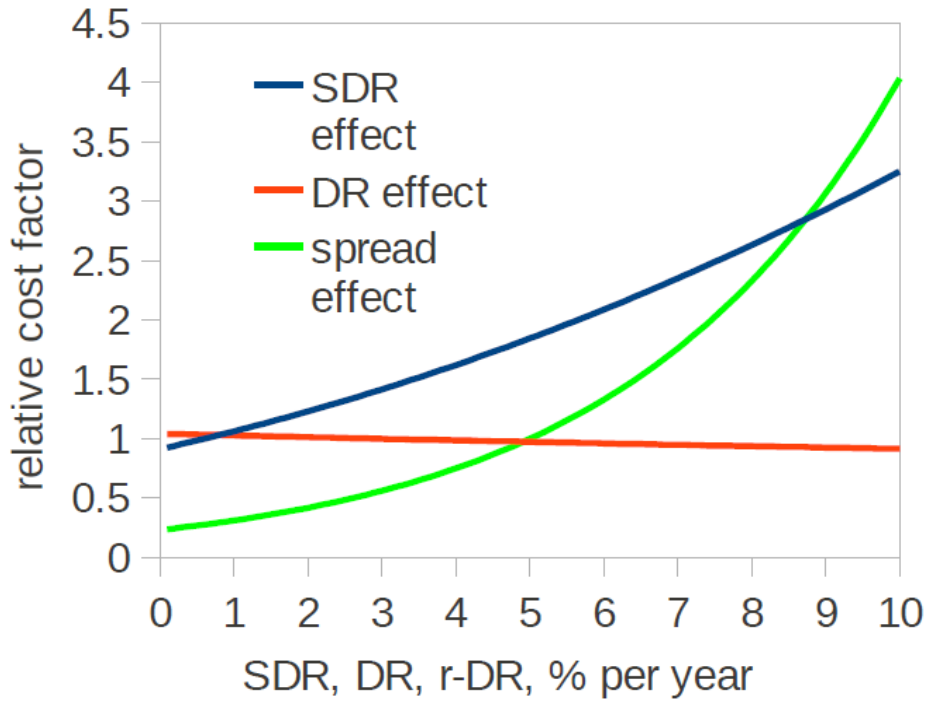

**Fig. 2.** The effect on cost of system degradation rate (blue), discount rate (red), and credit spread (green).

The credit risk specific to the borrower has the most profound effect of all factors. The cost can be decreased four-fold for a near risk-free borrower, whereas it doubles if the spread is increased from 5% to 7.5%. Specifically, for low-risk borrowers (spreads below about 4%), the response of the cost to relative change in spread is smaller than to relative change in conversion efficiency and insolation. Research into improvement of solar-to-electricity conversion efficiency is thereby re-given the priority it lost in Darling (2011).

The models we considered do not provide a complete cost-benefit analysis. Cost-benefit models will have to include the cash flow from electricity sales. That cash flow should be discounted using the risk-free rate plus premiums for the risks to fail to produce and to collect sales proceeds, which does not have to add up to the borrowing rate (e.g. due to different term structures or collaterization conditions for moneys owed by and to the company); using the latter could lead to an arbitrage opportunity whereby valuations will be depressed for riskier borrowers, and an entity with a better risk profile could buy the enterprise and enjoy higher valuations. Needless to say, the present argument about the appropriate discount rate is applicable to other industries as well (McIlveen-Wright, 2011).

In summary, we proposed a corrected description of financial factors influencing the cost of solar energy vs. recently proposed models (Azzopardi, 2011; Branker, 2011; Darling, 2011). Our model establishes the importance of the conversion efficiency (in contrast to Darling (2011)) and the need for a low credit risk of the enterprise (in contrast to both Azzopardi (2011) and Darling (2011)). We hope that the issues addressed in this Communication will help develop better quantitative models of economic performance of photovoltaic installations and avoid mis-allocation of research effort and of capital.

The author thanks A. Chablinskaia, Sr. Financial Planning Analyst, the city of Toronto, Canada, Dr. A. Ordine, Model Validation Group, Ontario Teachers' Pension Plan, Toronto, Canada, and Dr. D. Fedorets, Trading Quantitative Analyst, ING Bank, The Netherlands for proofreading the manuscript.